# Alter-Piezoresponse in Two-Dimensional Lieb-Lattice Altermagnets

Xilong Xu,[1] Li Yang*,[1,2]

[1]Department of Physics, Washington University in St. Louis, St. Louis, Missouri 63130, USA

[2]Institute of Materials Science and Engineering, Washington University in St. Louis, St. Louis, Missouri 63130, USA

* Email: lyang@physics.wustl.edu

## ABSTRACT

Altermagnetism, featuring alternating spin structures in reciprocal space, has sparked growing interest. Here, we predict novel real-space alternative piezomagnetic and piezoelectric responses in an emerging altermagnetic family of Lieb lattices, specifically transition-metal chalcogenides $M_2WS_4$ (M = Mn, Fe, Co). The unique $S_4T$ crystal-spin symmetry leads to distinct magnetic and electric responses depending on the direction of applied stress. When subjected to axial stress, they exhibit a giant piezomagnetic response, which is about one to two orders of magnitude larger than that of most piezomagnetic materials, while the residual $C_2$ symmetry suppresses the piezoelectric effect. In contrast, diagonal stress induces an imbalance of oppositely aligned electric dipole moments and a significant piezoelectric response, while in-plane mirror symmetry inhibits the piezomagnetic effect. This alternative piezoresponse offers an unprecedented opportunity to precisely control electric and magnetic properties independently, opening new avenues for altermagnetic materials in high-fidelity multifunctional memory and sensor applications.

Piezotronics, which regulates various physical order parameters through piezo-potential, has garnered significant attention and seen substantial growth in applications such as sensing, actuation, and energy harvesting.[1–8] Among the most extensively studied piezo-induced effects are the piezoelectric and piezomagnetic effects.[9–11] Discovered in 1880 by Jacques and Pierre Curie,[12] the piezoelectric effect enables signal transduction between mechanical and electrical domains: piezoelectric materials become electrically polarized under external strain and can be mechanically deformed by an applied voltage.[13–17] Similarly, piezomagnetic materials with net-zero spontaneous magnetization, which are much rare than the reported piezoelectricity, exhibit changes in magnetic moment under mechanical stress or produce magneto-strictive effects under an applied magnetic field.[18–21] Beyond these individual effects, single-phase materials that combine both piezoelectric and piezomagnetic effects and have the capability to selectively utilize one of them are even more intriguing. However, such material candidates are extremely limited.

Recently, rich and enhanced piezoelectric effects have been predicted and observed in numerous two-dimensional (2D) van der Waals (vdW) materials.[22–34] In contrast, the exploration of 2D piezomagnetic materials has been relatively limited. To date, only a few 2D materials have been reported to exhibit piezomagnetic effects. Among these, a particularly promising candidate pool comes from materials with a newly emergent magnetic order known as altermagnetic materials, which are characterized by alternative spin polarizations in reciprocal space.[35–48] Interesting piezoelectric and piezomagnetic responses have been reported in altermagnetic $V_2Se_2O$, $Cr_2S_2$ and Janus structures,[49–53] and the piezomagnetic response can be further enhanced via carrier doping.[49–51] However, how the unique crystal-spin symmetry of intrinsic altermagnets can lead to fundamentally unprecedented piezoresponses and how to selectively achieve a specific type of piezoresponses remain largely unexplored.

In this work, we predict a new class of 2D altermagnets, i.e., ternary transition-metal chalcogenides $M_2WS_4$ (M = Mn/Fe/Co) characterized by a $S_4T$ crystal-spin symmetry of Lieb lattices, can exhibit the uniquely alternative piezoresponses. When stress is applied along the axial direction, the material exhibits a giant piezomagnetic response with one to two orders of magnitude larger than those previously reported ones in altermagnets. However, the piezoelectric effect is forbidden for this axial strain due to the preservation of $C_2$ symmetry. Conversely, when the stress direction is switched to the diagonal directions, the material exhibits a significant piezoelectric response. Interestingly, the presence of in-plane mirror symmetry prevents the piezomagnetic effect under this diagonal strain. As a result, the system displays a "45-degree" alternatively piezomagnetic and piezoelectric responses. This phenomenon adds a new real-space alternative character to altermagnets and makes it possible to selectively control the electric and magnetic properties of materials, giving hope to novel sensors and multifunctional devices.

*Altermagnetic Lieb-Lattices:* The piezo-magnetoelectric effect discussed in this work is based on a family of 2D Lieb-lattice materials, $M_2WS_4$ (with M = Mn/Fe/Co), as shown in Fig. 1(a). Monolayer $M_2WS_4$ features a square lattice structure with a symmorphic space group *P-42m* (No. 111) and belongs to the $D_{2d}$ point group, characterized by the $S_{4z}$ and $C_{2x}$ symmetry operations. Its structure is composed of three atomic sublayers, where the magnetic Mn/Fe/Co and non-magnetic tungsten atoms form a central layer sandwiched between two sulfur atomic layers (see Figs. 1(a) and 1(b)). Furthermore, each metal atom is locally coordinated in a nearly perfect tetrahedral geometry, as depicted in Fig. 1(a). The dynamical stabilities of these structures have been verified through the phonon spectrum calculation (refer to Fig. S1), and the lattice parameters for this series are summarized in Table S3. Notably, a few these ternary transition-metal chalcogenides, such as $Cu_2WX_4$ (X = S/Se), $Cu_2MoS_4$, $Ag_2WS_4$, $MnAl_2S_4$, and $FeGa_2S_4$ have already been experimentally synthesized and exfoliated, significantly bolstering the feasibility of realizing the proposed materials in experiments.[54–61] Our calculated formation energies (see details in Section I in SI) indicate that the structure proposed in this work is more energetically favorable.

Our first-principles density functional theory (DFT) simulations (see details in Section II in SI) reveal that the altermagnetic order is the most stable in these Lieb-lattice magnets. In fact, a few members of this family, such as $Fe_2WS_4$, were recently proposed to be altermagnetic,[62] and this family of materials have recently attracted increasing research interest.[62,63] In the following, we will use monolayer $Mn_2WS_4$ as an example to demonstrate the predicted alter-piezoresponses. Other materials in this family exhibit similar properties, as summarized in Table I.

The band structure of $Mn_2WS_4$ is plotted in Fig. 1(c). It shows not only the characteristic flat bands around the Fermi level induced by Lieb lattices but also a distinct *d*-wave altermagnetic behavior. Specifically, the bands along the Γ–X–M and Γ–Y–M directions are energetically degenerate yet exhibit opposite spin orientations— a feature that originates from the antiferromagnetic (AFM) state combined with the material's unique crystal symmetry: all magnetic atom sites with the same spin exhibit a $C_2$ symmetry, whereas the sites with opposite spins lack translational or inversion symmetry. Instead, they are connected through the $S_{4z}$ symmetry, as illustrated in Fig. 1(d). This results in the formation of a combined spin symmetry group, $\{C_2\|S_4\}$, giving rise to the distinct *d*-wave alternating magnetic structure. However, the characteristic *d*-wave altermagnetism ensures that, at the high-symmetry X and Y points, the bands are spin-split, with opposite spins due to the $\{C_2\|M_{xy}\}$ spin symmetry.

This stable altermagnetic states in Lieb-lattice materials is based on the synergistic magnetic exchange mechanisms. $Mn_2WS_4$ exhibits a *d*-electron magnetism because of the localized 3*d* electrons from the manganese atom, and the tungsten and sulfur atoms contribute negligible magnetic moments. Based on the tetrahedral crystal-field theory, the five degenerate *d* orbitals of manganese atoms split into two subsets (i.e., the two-fold $e_g$ levels with a lower energy and three-fold $t_{2g}$ levels with higher energy), as illustrated in Fig. 1(e). Given the oxidation states, each manganese atom hosts approximately six *d* electrons, resulting in a magnetic moment of about 4 $\mu_B$. Considering the two nearest-neighbor manganese atoms, there are mainly two types of exchange coupling mechanisms. One is the direct exchange coupling. The other is the indirect exchange coupling mediated by the intermediate sulfur atoms, also named the super-exchange coupling. Based on the requirement of the orbital occupation of electrons, the direct exchange coupling results in a lower energy for the opposite-spin coupling. For the indirect exchange coupling, as illustrated in Fig. 1(e), the spin-up moment of the Mn(I) atom induces a spin polarization in the *p* orbitals of the intermediate sulfur atom. This polarization subsequently couples with the Mn(II) atom, further stabilizing the opposite spin alignment. This interaction relies on the bond angle between magnetic atoms and intermediate sulfur atoms. We find that the Mn–S–Mn bond angle is approximately 110°, which, according to the conventional Goodenough–Kanamori–Anderson (GKA) rules, would typically favor ferromagnetic coupling. However, since the *d* electrons of the Mn atoms are more than half-filled, the Pauli exclusion principle hinders ferromagnetic hopping, making the interaction more inclined toward an antiferromagnetic coupling. In conclusion, both coupling mechanisms collaboratively stabilize the altermagnetic ground state.

*Giant piezomagnetic effect:* As discussed earlier, the altermagnetic state—with its net zero magnetic moment—stems from the presence of the $M_{xy}$ or $S_4T$ symmetry. When these symmetries are disrupted, the balance of opposite spins is broken, resulting in a net magnetic moment. Uniaxial stress can offer a straightforward way to break these symmetries and trigger the piezomagnetic effect. For example, when uniaxial stress is applied along the [100] x-axial direction, the distance between the Mn(I) atom and its neighboring tungsten atoms (along with the connected sulfur atoms) is increased, altering the chemical environment and spin-charge distribution of this Mn(I) atom. In contrast, the Mn(II) atom, aligned with the [010] (y-axial) direction, undergoes minimal change, so its local chemical environment remains largely unaffected. This disparity in the local environments of Mn(I) and Mn(II) atoms leads to variations in their local magnetic moments, thereby generating a non-zero net magnetic moment. At this point, the system is no longer a perfect altermagnet but instead evolves into a ferrimagnetic state with a small magnetization. To demonstrate this idea, we have calculated the variation of the net magnetic moment under applied stress ranging from -0.5% to 0.5%, as shown in Fig. 2(a). The simulations reveal that applying either tensile or compressive 0.5% strain induces a magnetic moment of approximately 0.0017 $\mu_B$ per unit cell (u.c.). This induced moment is significantly larger than that found in intrinsic 2D materials like $V_2Se_2O$ and $V_2SeSO$,[49–51] and it is on par with that observed in $V_2Se_2O$, although the latter requires extrinsic hole doping.[49]

To further understand the microscopic picture of this piezomagnetic effect, we have computed the variation in Bader charges of the Mn(I) and Mn(II) atoms under applied stress, as plotted in Fig. 2(b). Under a +0.5% strain along the [100] direction, the Bader charge of the spin-up Mn(I) atom undergoes a noticeable decrease, with variations reaching up to $10^{-2}$ electrons, while that of the Mn(II) atom remains largely unaffected. According to Fig. 1(e), this induces an increase of the spin-up magnetic moment of the Mn(I) atom according to the Hund's rule,[64–66]. Consequently, the system exhibits a variable net magnetic moment, demonstrating the piezomagnetic effect.

This mechanism can be further used to explain the differences of piezomagnetic responses between the three members of our studied Lieb-lattice altermagnets: $Mn_2WS_4$, $Fe_2WS_4$ and $Co_2WS_4$. We found that under the same -0.5% strain, the Bader charge difference between the two magnetic atoms in a unit cell is 0.007$e$ for $Mn_2WS_4$, 0.004$e$ for $Fe_2WS_4$, and 0.008$e$ for $Co_2WS_4$, respectively. Despite the substantially different band gaps, this variation of Bader charge agrees with the trend of the piezomagnetic responses of our studied Lieb-lattice altermagnets. The essence of this difference lies in the varying $d$-electron counts, which lead to different orbital filling configurations. Based on this picture, applying stress along the [010] (y-axial) direction, however, shall produce the opposite outcome, which agrees with the red-color curve in Fig. 2(a). This directional dependence highlights the interplay between lattice distortion and the localized electronic structure, establishing a direct link between the piezomagnetic effect and uneven local charge distribution around the manganese atoms.

In the following, we quantitatively analyze the piezomagnetic response of this system. Considering that the easy magnetization axis of this material aligns along the z direction, the piezomagnetic effect can be expressed in the form of a third-order tensor: **M** = **q** • **ε**, where **M** represents the variation in the system's net magnetic moment, **q** denotes the piezomagnetic tensor, and **ε** represents the strain tensor. The **q** can be equated as:[67,68]

$$q_{ijk} = \frac{\partial M_i}{\partial \varepsilon_{jk}}. \tag{1}$$

Combining the 2D nature and the S$_4$T symmetry of these Lieb lattices, we find that there is only one independent component of the piezomagnetic tensor, which 0.34 μ$_B$/u.c. (see Section IV of SI for details),

However, this coefficient cannot be directly compared with experimental measurements, as the piezoelectric effects in practical applications are often evaluated using stress as a coefficient. Therefore, we convert to the form **M** = **Q** • **σ**, where **σ** represents the applied stress tensor, wherein **Q** can be defined as:[67]

$$q_{ijk} = \frac{\partial M_i}{\partial \varepsilon_{jk}} = \frac{\partial M_i}{\partial \sigma_{mn}} \frac{\partial \sigma_{mn}}{\partial \varepsilon_{jk}} = Q_{imn} C_{mnjk} \tag{2}$$

where $M_i$, $\varepsilon_{jk}$, and $\sigma_{jk}$ represent the piezomagnetic moments, strains, and stresses, respectively. $C_{mnjk}$ is the elastic stiffness tensor. For this purpose, we need to evaluate the Stiffness tensor and Young's modules (see section V of SI for details).[69] As plotted in Figs. 2(c), the in-plane Young's modulus is highly anisotropic. exhibiting a distinctly symmetric "cloverleaf" pattern that is attributable to the S$_4$ symmetry. Specifically, the softest mechanical response occurs along the diagonal [110] and [1$\bar{1}$0] directions—with a Young's modulus of 12.7 N/m while the stiffest response (47.4 N/m) are observed along the axial [100]/[010] directions. Interestingly, compared to other 2D materials like phosphorene (24–102 N/m),[70] MoS$_2$ (140–220 N/m),[71] and graphene (~340 N/m),[72] this material's relatively smaller Young's modulus allows it to undergo larger mechanical strains under lower stresses. Ultimately, we can obtain the piezomagnetic coefficient of these Lieb-lattice altermagnets. The results are summarized in Table I. For example, the piezomagnetic coefficient of monolayer Mn$_2$WS$_4$ is 3.6 x 10$^{-3}$ μ$_B$/Mn/N/m, and it is orders of magnitude larger than that reported in MnTe (3.9 x 10$^{-5}$

$\mu_B$/Mn/N/m) and MnF$_2$ (1.2 x 10$^{-4}$ $\mu_B$/Mn/N/m).[73,74]

It is noted that, if there is an altermagnetic Lieb lattice using chemically and structurally different building blocks (e.g., without octahedra), the piezomagnetic effect may still exist in principle as long as S$_4$T symmetry is broken and magnetic sublattices respond asymmetrically. However, the magnitude of the response may be smaller if spin–orbit coupling is weak or if magnetic sublattice asymmetry are reduced. Therefore, while the phenomenon is symmetry-allowed and conceptually general across Lieb-type magnetic lattices, the "giant" magnitude reported here is indeed amplified by the specific choice of elements and bonding environment.

Notably, when strain is applied along the axial [100] (x-axial) or [010] (y-axial) directions, the C$_2$ symmetry remains intact, as shown in Fig. 2(d). Consequently, the system preserves its in-plane axial rotational symmetry and inhibits the formation of a net electric dipole moment, the piezoelectric response.

*Piezoelectric effect:* When stress is applied along the diagonal directions ([110] or [1$\bar{1}$0]), the system exhibits a fundamentally different piezo-response. While the S$_4$T joint symmetry is broken, the M$_{xy}$ symmetry remains intact, ensuring that the system retains a zero net magnetic moment and thus eliminates the piezomagnetic effect. Thus, this system still maintains an altermagnetic state. Instead, we find that the breaking of the C$_2$ in-plane symmetry under diagonal stress leads to the emergence of a net electric dipole moment. Due to the in-plane mirror symmetry, this dipole moment is restricted to the z-direction, resulting in an out-of-plane piezoelectric effect. The calculated variation of the electrical polarization under applied diagonal stress is shown in Fig. 3(a): applying either tensile or compressive strain of 0.5% induces a net polarization of approximately +/- 0.114 pC/m.

In the following, we analyze the mechanism underlying this piezoelectric effect. In Fig. 3(b), the bottom sublayer sulfur atoms bridge metal atoms along the diagonal [110] direction, while the upper sublayer sulfur atoms bridge metal atoms along the orthogonal [1$\bar{1}$0] direction. Consequently, the strain along the [110] direction significantly alters the binding strength, atomic distances, and charge transfer between the middle sublayer of metal atoms and the bottom sulfur sublayer, affecting the electric dipole moment in the bottom part of the structure. Meanwhile, the upper sulfur sublayer, being orthogonal to the stress direction, remains relatively unaffected. This results in a piezoelectric response under the diagonal strain.

Another factor is from the material's positive Poisson ratio (Fig. S4): the lattice in the orthogonal [1$\bar{1}$0] direction compresses when tensile stress is applied along the diagonal [110] direction. This compression causes the dipole moment at the upper to respond oppositely to the bottom, amplifying the bonding variations between the top and bottom layers. This increasing asymmetry further breaks the $S_4$ symmetry, enhancing the net electric dipole moment.

To support this analysis, we calculated the variation in atomic distances between the middle metal atoms and the nearest upper and bottom sulfur atoms under diagonal [110] strain, as shown in Fig. 3(b). The results confirm that these the distances from top and bottom sublayers to the middle sublayer exhibit opposite trends under applied stress, contributing to an out-of-plane piezoelectric effect. We also obtained the piezoelectric responses of two other materials as listed in Table I. The differences in piezoelectric response between $Mn_2WS_4$, $Fe_2WS_4$ and $Co_2WS_4$ are from the variations of interaction strength between the metal atoms and sulfur atoms. This can be verified by examining the bond lengths: the S–W bond lengths in the three materials are 2.44 Å ($Mn_2WS_4$), 2.27 Å ($Fe_2WS_4$), and 2.25 Å ($Co_2WS_4$), respectively. Thus, the stronger the interaction, the weaker the piezoelectric response.

We also find that, when the strain direction is reversed to the [1$\bar{1}$0] direction, all results are flipped vertically because of the $S_4$ symmetry, which explains the emergence and tunability of the piezoelectric effect in these Lieb-lattice altermagnets.

Similarly, we can obtain the piezoelectric coefficient by calculating the changes in the electric dipole moment under applied stresses (see Section VI of SI for details). The calculated piezoelectric strain and stress coefficients are about 2.3 x $10^{-11}$ C/m and 1.76 pm/V, respectively. The piezoelectric stress coefficients are comparable with those found in the 2H phase of monolayer TMDCs, such as h-BN (1.26 pm/V),[24] $2H-MoS_2$ (2.97 pm/V),[76] $2H-WS_2$ (2.19 pm/V),[28] and bulk GaN (3.1 pm/V).[77] Importantly, the piezoelectric effect of our Lieb-lattice altermagnets is out-of-plane, distinguishing them from the above in-plane piezoelectric materials.

These piezomagnetic and piezoelectric effects are summarized in Table I. In summary, when stress is applied along axial directions [100]/[010], the enhanced piezomagnetic effect is observed, with the opposite responses along two axes. Conversely, stress applied along the diagonal directions [110]/[1$\bar{1}$0] induces a net electric dipole moment, showcasing a piezoelectric effect, again with the opposing responses between two diagonal directions. We also examined the impact of different

U values on the piezo effects and found that the piezomagnetic response exhibits reasonable variations while remaining within the same order of magnitude, and the piezoelectric effect is not changed much, as listed in supplementary materials Table S4 (see Section VII of SI for details). As illustrated in Fig. 4, these alternating piezomagnetic and piezoelectric responses, collectively termed alter-piezoresponses, originate from the unique $S_4T$ symmetry.

The exceptional piezo-magnetoelectric behavior unlocks significant potential for multifunctional applications to precisely control electric or magnetic properties via strain engineering. The reverse effects, such as electrostriction and magnetostriction, further enable the detection of electric and magnetic fields via strain responses, such as dual-mode sensing structures capable of detecting stress with special resolution through electrical or magnetic signals. Finally, when subjected to periodic tensile or compressive stress, the material exhibits a periodic electromagnetic response, with electric and magnetic signals alternating at a period of $\pi$ and maintaining a quarter-period ($\pi/4$) phase difference. This dynamic response introduces novel possibilities for studying alternating electromagnetic phenomena and developing advanced materials and devices.

## AUTHOR INFORMATION

Corresponding Author:

**Li Yang** - Department of Physics and Institute of Materials Science and Engineering, Washington University in St. Louis, St. Louis, Missouri 63130, USA

Email: lyang@physics.wustl.edu

Author:

**Xilong Xu** - Department of Physics, Washington University in St. Louis, St. Louis, Missouri 63130, USA

Email: xilong@wustl.edu

**Notes**

The authors declare no competing financial interest.

## ASSOCIATED CONTENT

Supporting Information

The Supporting Information is available free of charge at URL.

We provide the phonon spectrum, lattice constant and the calculation details of piezomagnetic and piezoelectric coefficient, Young's modulus and Poisson's ratio. We also provide the effect on Piezo responses of different U values.

## ACKNOWLEDGMENTS

X.X. is supported by the National Science Foundation (NSF) Designing Materials to Revolutionize and Engineer our Future (DMREF) DMR-2118779. L.Y. is supported by NSF DMR-2124934. The simulation used Anvil at Purdue University through allocation DMR100005 from the Advanced Cyberinfrastructure Coordination Ecosystem: Services & Support (ACCESS) program, which is supported by National Science Foundation grants #2138259, #2138286, #2138307, #2137603, and #2138296.

**Figures:**

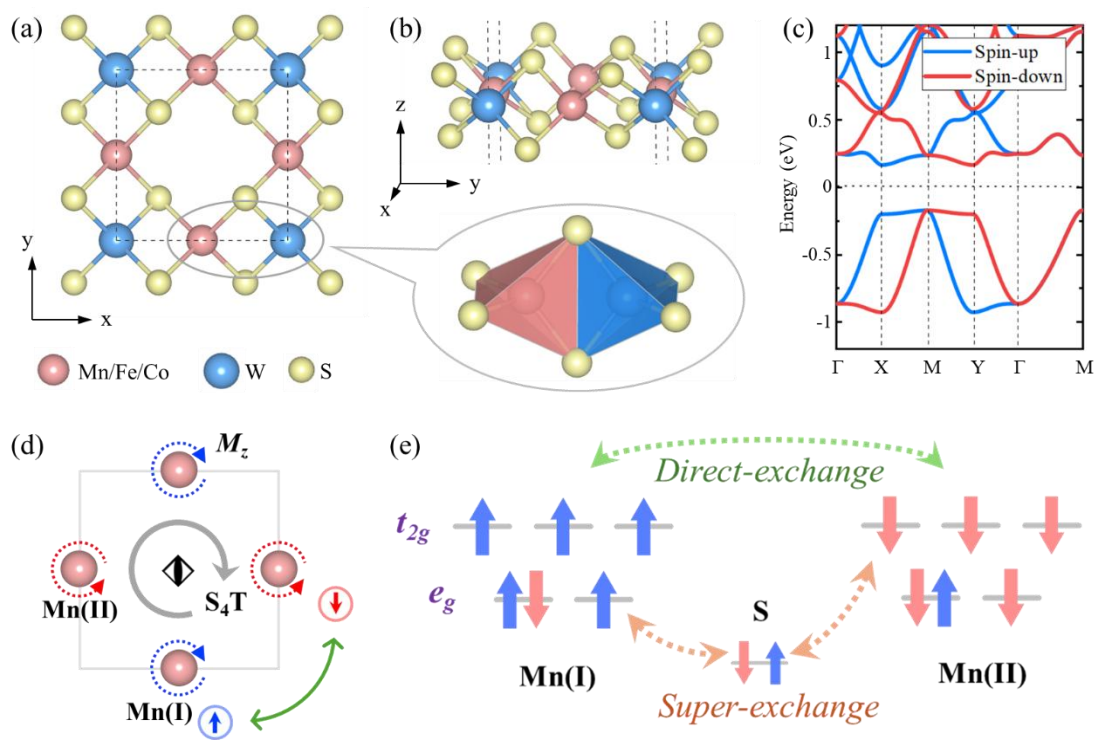

**Fig. 1**. (a) and (b) top view and side view of crystal structures of monolayer $M_2WX_4$. (c) Band structures of $Mn_2WS_4$. (d) The schematic of altermagnetism in $Mn_2WS_4$. (e) The schematic of intralayer exchange coupling between nearest Mn atoms.

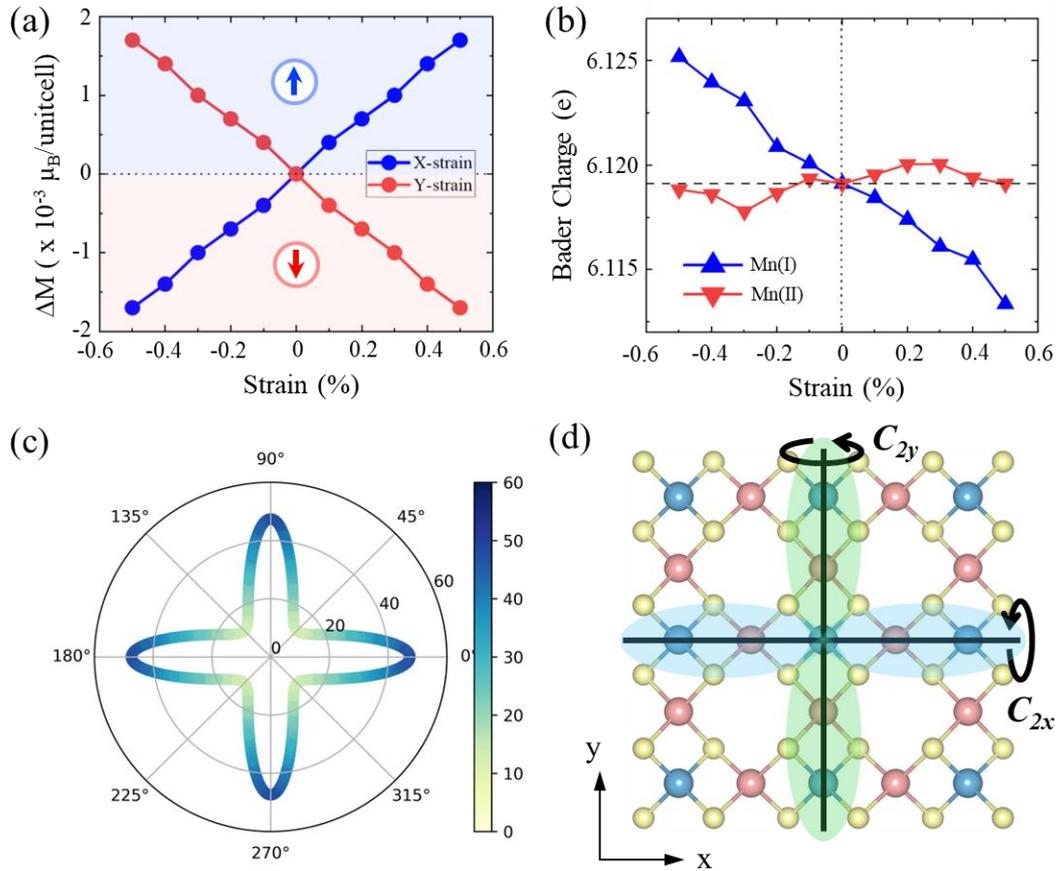

**Fig. 2.** (a) Variation in the system's net magnetic moment of $Mn_2WS_4$ under applied stress ranging from -0.5% to 0.5% along [100] or [010] directions. (b) Variation in Bader charges of the $Mn_1$ and $Mn_2$ atoms of $Mn_2WS_4$ under applied stress ranging from -0.5% to 0.5% along [100] directions. (c) Young's modulus of $Mn_2WS_4$ as a function of angle θ. θ = 0° corresponds to [100] direction. (d) The schematic of strained $Mn_2WS_4$ and residual symmetry along the [100] or [010] direction.

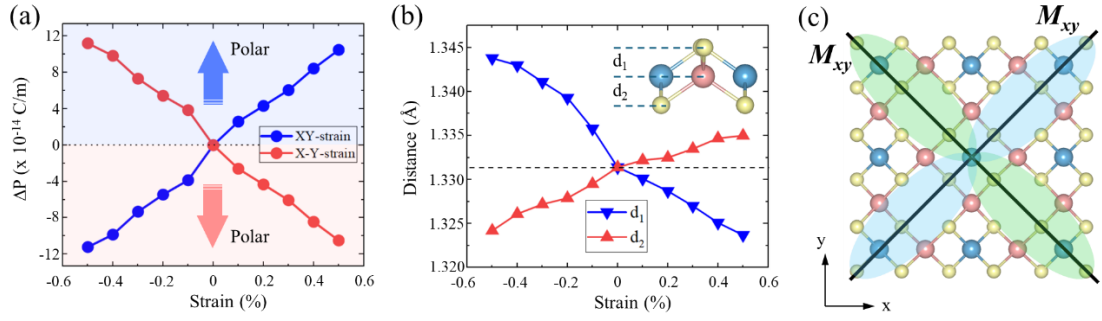

**Fig. 3**. (a) Variation in the system's net polarization of $Mn_2WS_4$ under applied stress ranging from -0.5% to 0.5% along [110] or [1$\bar{1}$0] directions. (b) Variation in the distances between the $Mn_1$ and $Mn_2$ atoms of $Mn_2WS_4$ under applied stress ranging from -0.5% to 0.5% along [110] directions. (c) The schematic of strained $Mn_2WS_4$ and residual symmetry along [110] or [1$\bar{1}$0] directions.

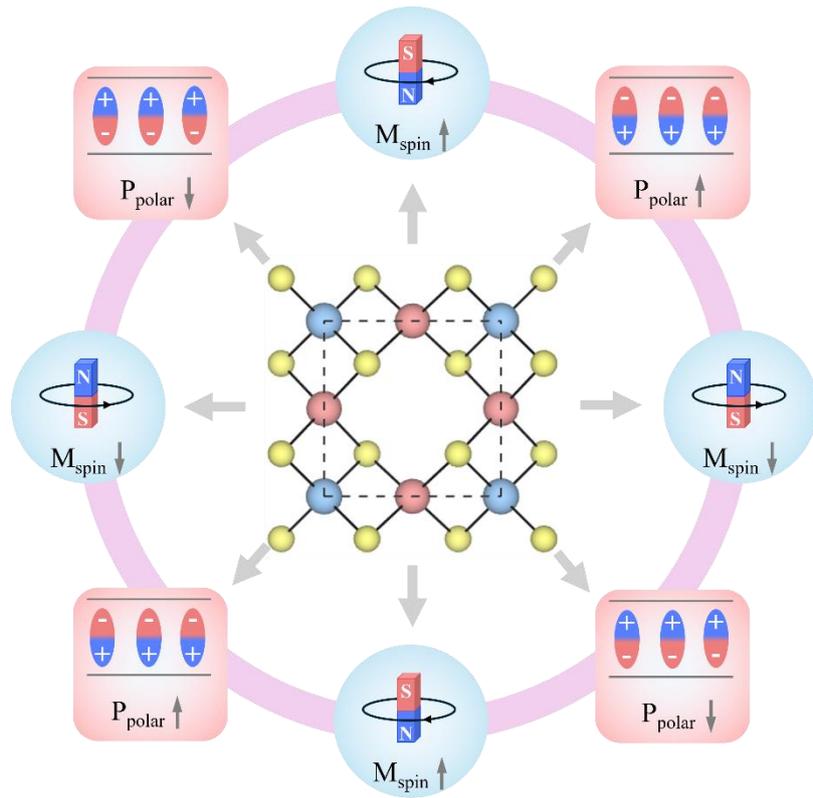

**Fig. 4**. The schematic of alter-piezoresponses in $Mn_2WS_4$.

**Table I**. Summary of calculated piezomagnetic and piezoelectric strain coefficients of Lieb-lattice altermagnets and other materials.

|  | Piezomagnetic ($\mu_B$/unit cell) | | piezoelectric (x $10^{-10}$ C/m) | |
| --- | --- | --- | --- | --- |
|  | Axial | Diagonal | Axial | Diagonal |
| **$Mn_2WS_4$** | 0.34 | 0 | 0 | 0.23 |
| **$Fe_2WS_4$** | 0.016 | 0 | 0 | 0.11 |
| **$Co_2WS_4$** | 0.50 | 0 | 0 | 0.08 |
| $V_2Se_2O$ | <0.01 | 0 | null | |
| $Cr_2S_2$ | <0.01 | 0 | null | |
| MnTe (bulk) | ~ 0.0027 | | null | |
| $MnF_2$ (bulk) | ~ 0.017 | | null | |
| BN | null | | 1.38 | |
| $MoS_2$ | null | | 3.64 | |
| GeS | null | | 4.6/-10.1 | |

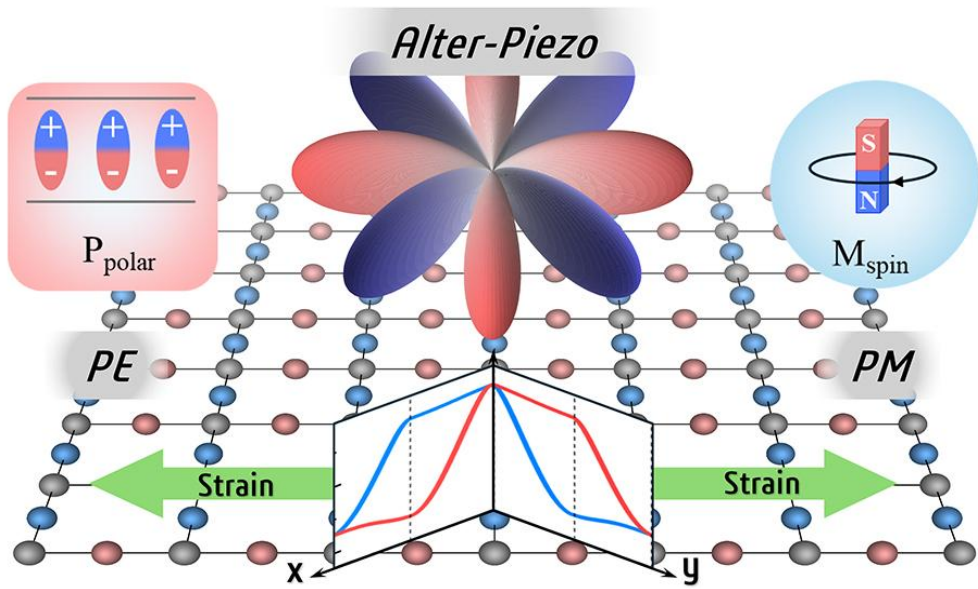

**TOC Graphic**